\documentclass[a4paper,12pt,epsf]{article}
\usepackage{amsmath}
\usepackage{amssymb}
\usepackage{amsfonts}
\usepackage{graphicx}
\usepackage{booktabs}
\usepackage{amsmath,amssymb,slashed}

\renewcommand{\l}{\left}

\newcommand{\q}{\partial}
\renewcommand{\r}{\right}

\newcommand{\beq}{\begin{eqnarray}}
\newcommand{\eeq}{\end{eqnarray}}

\newcommand{\GeV}{\,{\rm GeV}}
\newcommand{\TeV}{\,{\rm TeV}}
\newcommand{\nn}{\nonumber \\}

\newcommand{\SU}{\mathop{\rm SU}}
\newcommand{\U}{\mathop{\rm U}}

\newcommand{\s}[1]{_{\mathrm{#1}}}
\newcommand{\MET}{\slashed{E}_T}
\newcommand{\Order}{\mathcal{O}}
\newcommand{\sgn}{\mathop{\mathrm{sgn}}}

\begin{document}


\baselineskip 0.7cm

\begin{titlepage}

\begin{flushright}
IPMU--14--0186
\end{flushright}

\vskip 1.35cm
\begin{center}
{\large \bf\boldmath
 CP-safe Gravity Mediation and Muon $g-2$
}
\vskip 1.2cm
Sho Iwamoto, Tsutomu T.~Yanagida and Norimi Yokozaki
\vskip 0.4cm

{\it  Kavli IPMU (WPI), the University of Tokyo, Kashiwa, Chiba 277--8568, Japan}

\vskip 1.5cm

\abstract{
We propose a CP-safe gravity mediation model, where the phases of the Higgs B~parameter, scalar trilinear couplings and gaugino mass parameters are all aligned.
Since all dangerous CP violating phases are suppressed, we are now safe to consider low-energy SUSY scenarios.
As an application, we consider a gravity mediation model explaining the observed muon $g-2$ anomaly. 
The CP-safe property originates in two simple assumptions: SUSY breaking in the K\"ahler potential and a shift symmetry of a SUSY breaking field $Z$. As a result of the shift symmetry, the imaginary part of $Z$ behaves as a QCD axion, leading to an intriguing possibility: the strong CP problem in QCD and the SUSY CP problem are solved simultaneously.
}

\end{center}
\end{titlepage}

\setcounter{page}{2}

\section{Introduction}
Large CP violation is a generic problem in the supersymmetric (SUSY) standard model (SSM).
In particular, this problem becomes extremely serious when sleptons, binos and winos are as light as $\Order(100)\GeV$. 
In such cases,  SUSY contributions to the electric dipole moment (EDM) of the electron usually exceed the experimental bound~\cite{edm_acme2014} by many orders of magnitude, due to CP violation in the SUSY breaking sector~\cite{edm_susy}.
On the other hand, the observed muon $g-2$ anomaly~\cite{g-2_bnl2010} suggests such light sleptons, bino and wino; the experimental value of the muon $g-2$ can be explained by the contributions from the bino, wino and sleptons of masses $\Order(100)$\,GeV~\cite{SUSY_gminus2}.
Moreover, models with such light SUSY particles are fascinating possibilities for the ILC.
Therefore, it is important to construct a mechanism to suppress CP violating phases of the SSM.
The SSM has many sources of the CP violating phases: soft SUSY breaking masses of sfermions, gaugino mass parameters, scalar trilinear couplings and the Higgs B-term.
The first source, linked to the flavor structure, can be eliminated by assuming that the soft SUSY breaking masses are universal or zero at the high-energy scale.
Thus, we restrict our discussion to the other sources.

In this paper, we propose a CP-safe framework of the gravity mediation in the SSM, where all of the relevant phases are aligned. Consequently, the dangerous CP violating phases are suppressed enough, and hence low-energy SUSY scenarios become attractive.
Encouraged by this theoretical proposal, we construct a model consistent with the observed anomaly of the muon $g-2$. We also show how to test this model at the LHC and linear colliders.

\section{CP problem in the SSM}

CP violation in soft SUSY breaking terms is serious obstacle for the low-energy SUSY. Apart from CP violation in mass matrices of sfermions, there are dangerous CP violating phases in the Higgs sector, gaugino masses and trilinear couplings of sfermions. In fact, those phases are severely constrained from the EDM experiments if the masses of the SUSY particles are $\Order(100\text{--}1000)\GeV$.\footnote{
Masses of non-colored SUSY particles are still allowed to be $\Order(100)$ GeV.
}

The Lagrangian of the relevant part is given by
\begin{equation}
\begin{split}
  \mathcal{L} &= \int d^2 \theta \Bigr[ (\mu - B_{\mu} \theta^2) H_u H_d  - \frac{1}{2} M_i \theta^2 \lambda_i \lambda_i
\\ &\quad
 + (1-(A_e)_{ij} \theta^2) (Y_e)_{ij} L_i H_d E_j^c + (1-(A_d)_{ij} \theta^2) (Y_d)_{ij} Q_i H_d D_j^c
\\ &\quad
+ (1-(A_u)_{ij} \theta^2) (Y_u)_{ij} Q_i H_u U_j^c   + \text{h.c.} \Bigl],
\end{split}
\end{equation}
where $\lambda_1$, $\lambda_2$ and $\lambda_3$ denote bino, wino and gluino, respectively. 
CP violation arises from the gaugino mass parameters $M_i$, the trilinear couplings $A_{e,d,u}$, the Higgsino mass parameter $\mu$ and the Higgs B-term $B_\mu$.
Physical CP violating phases are given by combinations of the phases of $B_\mu, \mu, M_i, A_{e,d,u}$ as%
\footnote{
This fact can be understood by taking the convention in which $\mu$ and $B_\mu$ are real, where the parameters are rotated as
\begin{align*}
& \mu, B_\mu \to |\mu|, |B_\mu|,
&
& M_i \to M_i \exp(i \arg (\mu B_\mu^*)) = |M_i| e^{i \theta_i},
&
& A_f \to A_f \exp(i \arg(\mu B_\mu^*)) = |A_f| e^{i\theta_{A_f}}.
\end{align*}
}
\begin{align}
 \theta_i&=\arg(M_i (\mu B_\mu^*)),
&
\theta_{A_{e,d,u}}&=\arg(A_{e,d,u} (\mu B_\mu^*)), \label{eq:angles}
\end{align}
where we neglect small CP violating phases of the Higgs vacuum expectation values (VEVs).

In the following, we focus on the electron EDM.
It is generated by the phases $\theta_1$, $\theta_2$ and $\theta_{A_e}$, and constrained as
\begin{equation}
 |d_e|  < 8.7 \times10^{-29} e\,{\rm cm},
\end{equation}
by ACME experiment~\cite{edm_acme2014}.
It generally gives the most stringent constraint for scenarios with light sleptons, and is especially severe for models with $|\mu|\tan\beta\gg M_1$, $M_2$, which we will utilize in the rest of this letter.
The large $|\mu|\tan\beta$ enhances the SUSY contribution to the muon $g-2$, and simultaneously, to the electron EDM.
Taking the common mass scale $m_{\rm SUSY}$ for $|M_1|$, $|M_2|$ and selectron masses, the contribution due to $\theta_1$ is approximated as~\cite{Pokorski:1999hz}
\begin{equation}
 \left(|d_e|/e\right)_{\theta_1} \approx \frac{m_e\alpha\s{EM}}{48\pi\cos^2\theta_w}\frac{|\mu|\tan\beta}{m\s{SUSY}^3}|\theta_1|.
\end{equation}
Assuming no cancellation with contributions from the other phases, the experimental bound gives a constraint
\begin{equation}
 \theta_1 \lesssim 1.2\times10^{-4}\l(\frac{m_{\rm SUSY}}{300\GeV}\r)^3 \l(\frac{3\TeV\times10}{|\mu|\tan\beta}\r). \label{eq:edm_const}
\end{equation}
Thus the phases of $B_\mu \mu^*$ and $M_1$ must be very accurately aligned at the scale of electroweak symmetry breaking (EWSB), or very small (see Eq.~(\ref{eq:angles})).

It should be noted that the phase of $B_\mu \mu^*$ depends on renormalization scale and is changed by those of the trilinear couplings $A_{u,d,e}$ and the gluino mass $M_3$. The relevant renormalization group equations for $B_\mu$ are~\cite{Martin:1993zk}
\begin{align}
 (16\pi^2) \frac{d (B_\mu/ \mu)}{d t}
   &\simeq
   6 (A_u)_{33} |(Y_u)_{33}|^2 + 6 (A_d)_{33} |(Y_d)_{33}|^2
\nn&\quad
   +  2 (A_e)_{33} |(Y_e)_{33}|^2
 + 6g_2^2 M_2 + \frac{6}{5}g_1^2 M_1,
\nn
 (16\pi^2) \frac{d (A_{u,d})_{33}}{d t} &\simeq   \frac{32}{3} g_3^2 M_3 + \dots,  \label{eq:rges}
\end{align}
where we neglect small Yukawa couplings. Here, $g_i$ are the gauge couplings constants for $\SU(3)$, $\SU(2)$ and $\U(1)$ gauge groups.
As seen from Eq.~(\ref{eq:rges}), even if the phases of $B_\mu \mu^*$ and $M_1$
are aligned at the high-energy such as $\simeq 10^{16}$ GeV, 
the resultant CP phase $\theta_1$ at the low-energy become $\mathcal{O}(1)$ due to the phases of $A_{u,d,e}$ and $M_3$.
Consequently, the constraints Eq.~(\ref{eq:edm_const}) are satisfied only when all the relevant phases, i.e., those of $B_\mu\mu^*$, $M_i$ and $A_{u,d,e}$, are precisely aligned or very small at the high-energy scale as long as there are no miraculous cancellations.

However, it is challenging in general to align all of the phases at the high-energy scale. To see this, we consider the following simple K\"ahler potential and the superpotential, so-called the Polonyi model~\cite{Polonyi:1977pj}:
\begin{align}
 K&=Z^* Z + (Q_{\rm SM}^*)_i (Q_{\rm SM})_i,
&
W &= \mu H_u H_d + W_{\rm Yukawa} + \mu_Z^2 Z + \mathcal{C},
\end{align}
where $Z$ is the SUSY breaking field and $\mathcal{C}$ is a complex constant.
Gaugino masses are generated from couplings between $Z$ and field strength superfields,
\begin{equation}
 \mathcal L \supset
 \int d^2 \theta \left(\frac{1}{4g_i^2} + k_i \frac{Z}{M_P} \right) W^{i \alpha} W_{\alpha}^i  + \text{h.c.},
\label{eq:gaugino_mass_before_rotation}
\end{equation}
where $M_P$ is the reduced Planck mass. 
To focus on CP violation in the SUSY breaking sector, we assume that the phases of $k_i$ are universal as $k_i = |k_i| e^{i \theta}$ in this section. For instance, the $\SU(5)$ GUT model satisfies this condition. 
Then, by a rotation of $Z$, the coefficients $k_i$ can be taken to be real.\footnote{
By using $R$-rotation, the constant $\mathcal{C}$ can be taken as real. However, to clarify the point, we leave $\mathcal{C}$ as a complex parameter.
}
In this basis, the scalar potential is given by\footnote{The replacement of the coupling, $1/(4g_i^2) + {\rm Re}(k_i z)\to 1/(4g_i^2)$, has been done. The gauge kinetic functions are taken to be canonical. }
\begin{equation}
  V
 =
 e^{K/M_P^2} 
 \Biggl[ 
 \left|\mu^2_Z +  Z^* \frac{W}{M_P^2} \right|^2 +
 \left|\frac{\q W}{\q (Q_{\rm SM})_i} + (Q_{\rm SM}^*)_i \frac{W}{M_P^2} \right|^2 
 -3\frac{|W|^2}{M_P^2}
 \Biggr],
\end{equation}
and the gaugino mass term of Eq.~(\ref{eq:gaugino_mass_before_rotation}) is now written as
\begin{equation}
 \mathcal L \supset
  k_i g_i^2 \frac{\l<F_Z \r>}{M_P} \lambda^i \lambda^i + \text{h.c.};
\end{equation}
the gaugino masses are proportional to $\left<F_Z\right>$:
\begin{equation}
 M_i = {2 k_i g_i^2} \frac{\l<F_Z\r>}{M_P}. \label{eq:mgaugino}
\end{equation}

SUSY breaking occurs by the nonzero F-term of $Z$:
\begin{equation}
\begin{split}
  \left< F_Z \right> &= - e^{K/(2M_P^2)}   \left( \frac{\q W^*}{\q Z^*} + \frac{\q K}{\q Z^*} \frac{W^*}{M_P^2} \right)
\\ &= - e^{ |z|^2/(2M_P^2)}   \left( \mu_Z^{2\,*}  + \frac{(\mu_Z^{2\,*} |z|^2 + \mathcal{C}^* z)}{M_P^2} \right). \label{eq:fzp}
\end{split}
\end{equation}
Here, $z$ is the VEV of $Z$, which is found to be 
\begin{align}
  z&\equiv \l<Z\r>=(\sqrt{3}-1) e^{i\theta_Z} M_P, 
&
\theta_Z&= \arg(\mathcal{C})-\arg(\mu_Z^2) \label{eq:vevz}
\end{align}
for the stable SUSY breaking vacuum with vanishing cosmological constant $V=0$. 
As a result, the phase of $\l<F_Z\r>$ turns out to be $\arg(\l<F_Z\r>)= \pi- \arg(\mu_Z^2)$.
The constant term is found to be
\begin{equation}
 |\mathcal{C}|=(2-\sqrt{3})|\mu_Z^2| M_P. \label{eq:const}
\end{equation}

The Higgs B-term and the A-terms are generated from 
\begin{equation}
  V 
\ni -e^{K/(2M_P^2)} \l<F_Z\r> \l<\frac{\q K}{\q Z} \r> \frac{W_{\rm vis}}{M_P^2}
  + e^{K/M_P^2} \l[ \frac{\q W_{\rm vis}}{\q (Q_{SM})_i}(Q_{SM})_i - 3 W_{\rm vis} \r]\frac{\left<W^*\right>}{M_P^2}, \label{eq:b_a_terms}
\end{equation}
where $W_{\rm vis}=\mu H_u H_d + W_{\rm Yukawa}$ is the superpotential of the visible sector.
Using Eqs.~(\ref{eq:fzp})--(\ref{eq:const}), we have 
\begin{align}
 B_\mu &= (2/\sqrt{3}-1) e^{-i \arg(\mathcal{C})} \frac{|\l<F_Z\r>|}{M_P} \mu,
&
 A_{ijk} &= (\sqrt{3}-1) e^{-i \arg(\mathcal{C})} \frac{|\l<F_Z \r>|}{M_P	}.
\end{align}
In this basis, $B_\mu \mu^*$ and $A_{ijk}$ have the same phases as $\mathcal C^*$, while the phases of the gaugino masses are given by $\arg(M_i)=\arg(F_Z)=\pi-\arg(\mu_Z^2)$.\footnote{
Of course the superpotential can be more generic; $\arg(B_\mu \mu^*)=\arg(A_{ijk})$ is not satisfied in the general case (see Appendix A).}
As  $\arg(\mathcal C)\neq\arg(\mu_Z^2)$ in general, the phases of $B_\mu \mu^*$ ($A_{ijk}$) and $M_i$ are different at the high energy.
Thus the resultant CP violating phases become too large at the electroweak scale, and the EDM exceeds the constraints of Eq.~(\ref{eq:edm_const}) unless the sleptons are as heavy as $\Order(10)\TeV$.


%
%
%
%

\section{CP-safe gravity mediation}
The Polonyi model has a large CP violation unless $\arg(\mu_Z^2)=\arg(\mathcal{C})$.
As one can see from Eqs.~(\ref{eq:mgaugino}) and (\ref{eq:b_a_terms}), the phases of the B-terms and A-terms are determined by those of $\l<F_Z\r> (\q K)/(\q Z)$ and $\l<W^*\r>$, 
while those of the gaugino masses are determined by the phase of $\l<F_Z\r>$. 
When the phases of $\mu_Z^2$ and $\mathcal{C}$ are aligned, $\l<Z\r>$ becomes real (see Eq.~(\ref{eq:vevz})). Then, $(\q K)/(\q Z)$ is real and $\arg(\l<F_Z\r>)= \arg(\left<W^*\right>)$ is satisfied (see Eq.~(\ref{eq:fzp})). As a result, the phases of the B-term, A-terms and gaugino masses are all aligned.  However, there is no reason why the phases of $\mu_Z^2$ and $\mathcal{C}$ are aligned.

We now propose a generic gravity mediation model where the 
CP-safe conditions, (a)\,$(\q K)/(\q Z) \in {\mathbb{R}}$ and (b)\,$\arg(\l<F_Z\r>)= \arg(\left<W^*\right>)$, are  automatically satisfied.
%
%
We adopt the gravitational SUSY breaking scenario~\cite{Izawa:2010ym} (see also Appendix B for details), where SUSY is broken in  the K\"ahler potential rather than in the superpotential:
SUSY is broken with a constant superpotential.\footnote{%
Considering that the super-Weyl--K\"ahler transformation is anomalous, we adapt the constant superpotential from the beginning.
}
Furthermore, we assume a shift symmetry of $Z$, which guarantees $(\q K)/(\q Z)$ to be real as shown below.
Consequently, the conditions (a) and (b) are automatically satisfied (see Eq.~(\ref{eq:fzp})).\footnote{
In Ref.~\cite{choisan}, CP violation in SUSY breaking is discussed with a Lagrangian motivated by the string theory. In their setup, parameters in a superpotential of SUSY breaking moduli can be taken to be real by approximate shift symmetries of moduli, resulting in no CP violation in the gaugino mass parameters and trilinear couplings. However, 
it is not clear if the CP violating phase of the Higgs B parameter is sufficiently aligned  with the other phases in their approaches; for our purpose to explain the muon $g-2$ anomaly, we need the CP violating phase suppressed as tightly as $\Order(10^{-4})$ level, because SUSY contribution to the electron EDM is enhanced as well as that to the muon $g-2$.
}

To demonstrate the above CP-safe gravity mediation, we consider the following Lagrangian:~\footnote{
As shown in Appendix D, the sequestered form of the K\"ahler potential is also consistent with a CP-safe gravity mediation.
}
\begin{align}
K &=  s(x) + (Q_{\rm SM}^*)_i (Q_{\rm SM})_i,
  &
W &= \mu H_u H_d + W_{\rm Yukawa}+ \mathcal{C}, \label{eq:cpsafe_lag}
\end{align}
where $s(x)$ is a real function of $x$, and $x=(Z+Z^*)$.
The Lagrangian has a shift symmetry
\begin{equation}
 Z \to Z + i \,\mathcal{R},
\end{equation}
where $\mathcal{R}$ is a real constant.
Notice that the R-charge of $(H_u H_d)$ is 2, which forbids regeneration of $\mu$ and $B_\mu$ from $K \ni c_n(Z+Z^\dag)^n (H_u H_d)+\text{h.c.}$
This is very important, since otherwise we may have a CP violating phase in the Higgs B-term.
Note that $K \ni d_n C^*(Z+Z^\dag)^n (H_u H_d) + \text{h.c.}$ allowed by the symmetry is not dangerous, since the generated $\mu$-term and B-term are suppressed  by $\mathcal{O}(m_{3/2}/M_P)$.

The scalar potential is then given by
\begin{equation}
  V = e^{K/M_P^2} \left[ \frac{1}{M_P^2} \left(\frac{\q s}{\q x} \right)^2 \left(\frac{\q^2 s}{\q x^2} \right)^{-1} -3  
 \right] \frac{|W|^2}{M_P^2} 
 + e^{K/M_P^2} \left| 
 \frac{\q W}{\q (Q_{\rm SM})_i} + \frac{\q K}{\q (Q_{\rm SM})_i} \frac{W}{M_P^2} 
 \right|^2 . \label{eq:cpsafe_grav}
\end{equation}
The SUSY breaking F-term of the hidden sector field $Z$ is proportional to the constant term
\begin{equation}
 \left< F_Z \right>= - e^{s/(2M_P^2)} \left(\frac{\q s}{\q x} \right) \left(\frac{\q^2 s}{\q x^2} \right)^{-1} \frac{\mathcal{C}^*}{M_P^2}. \label{eq:fz}
\end{equation}
Note that the shift symmetry of $Z$ guarantees $(\q s)/(\q x) \in \mathbb{R}$, and the phases of $\l<F_Z\r>$ is the same as that of $\l<W^*\r>=\mathcal{C}^*$. Now, it is clear that the CP-safe conditions (a) and (b) are satisfied.

The condition for the vanishing cosmological constant, $V=0$, is satisfied with appropriate choice of the K\"ahler potential rather than the tuning of the constant $\mathcal{C}$;
\begin{equation}
 \frac{1}{M_P^2}  \left(\frac{\q s}{\q x} \right)^2 \left(\frac{\q^2 s}{\q x^2} \right)^{-1} \Bigg|_{x=\left<x\right>} = 3, \label{eq:vanish}
\end{equation}
where the VEV $\left< x \right>$ is determined by a stationary condition,
\begin{equation}
 \frac{\q }{\q x} \left[ \left(\frac{\q s}{\q x} \right)^2 \left(\frac{\q^2 s}{\q x^2} \right)^{-1}  
\right] =0. \label{eq:min}
\end{equation}
A concrete example model which satisfies Eq.~(\ref{eq:vanish}) and Eq.~(\ref{eq:min}) with a stable minimum is shown in Appendix C. The constant $\mathcal{C}$ can be written using the gravitino mass $m_{3/2}$ as
\begin{equation}
|\mathcal{C}|^2 = e^{-s/(M_P^2)} m_{3/2}^2 M_P^4 .
\end{equation}

The Higgs B-term and A-term arises from the second term of Eq.~(\ref{eq:cpsafe_grav}) and they are proportional to the constant $\mathcal{C}^*$ as\footnote{
The superpotential in the visible sector is rescaled as $W_{\rm vis} \to e^{-K/(2M_P^2)} W_{\rm vis}$, such that $V$ reproduces the result in the global limit, $V \ni |({\q W_{\rm vis}})/({\q Q_i})|^2$.
}
\begin{align}
 B_\mu &=  e^{s/(2M_P^2)} \frac{2\mathcal{C}^*}{M_P^2} \mu,
&
 A_{ijk}&=  e^{s/(2M_P^2)} \frac{3\mathcal{C}^*}{M_P^2}.
\end{align}
Notice that the phases of $B_\mu \mu^*$ and $A_{ijk}$ are aligned with that of $\l< F_Z \r>$.
By using Eq.~(\ref{eq:fz}), they are rewritten as
\begin{align}
 B_\mu &=   - \frac{2}{3} \l(\frac{\q s}{\q x} \r) \frac{\l< F_Z \r>}{M_P^2} \mu , &
A_{ijk}&= - \l(\frac{\q s}{\q x} \r) \frac{\l< F_Z \r>}{M_P^2}. \label{eq:b_and_a}
\end{align}
Scalar masses are the same as the gravitino mass:
\begin{equation}
 {m}_0^2 =  \frac{|\l<F_Z\r>|^2}{9 M_P^4} \l(\frac{\q s}{\q x}\r)^2 = m_{3/2}^2. 
 \label{eq:scalar_soft_mass}
\end{equation}

Finally, let us consider the gaugino masses, which arise from the couplings between $Z$ and field strength superfields $W_i$:
\begin{equation}
 \int d^2 \theta \left( \frac{1}{4g_i^2} + k_i \frac{Z}{M_P} \right) W_{\alpha}^i W^{\alpha \, i}. \label{eq:z_gauge}
\end{equation}
However, the required $Z W_{\alpha}^i W^{\alpha\, i}$ terms violate the shift symmetry, and hence the terms must vanish, $k_i=0$. However, we consider the $Z W_{\alpha}^i W^{\alpha\, i}$ terms are generated by gauge anomalies of the shift symmetry, 
  resulting in $k_i \in \mathbb{R}$. The constants $k_i$ depend on unknown high-energy physics, and hence, we take them as free parameters in this paper.
%

Now, we see that all relevant phases are aligned as $\arg(B_\mu \mu^*)=\arg(M_i)=\arg(A_{ijk})$ and that the SUSY contributions to the EDM are successfully suppressed. We call this a CP-safe gravity mediation. Note that this feature is not affected by the anomaly mediation effect~\cite{amsb_org}. This is because  the contributions from the anomaly mediation to $A_{ijk}$, $B_\mu/\mu$ and gaugino masses are also aligned with $\mathcal{C}^* (= \arg(F_Z))$~\cite{amsb_general}, and any additional CP violating phases are not introduced by the anomaly mediation.


%
%
%
%
\section{Application: SUSY solving muon $g-2$ anomaly}

We have seen that the model we proposed, the CP-safe gravity mediation, provides SUSY breaking without CP violation. This feature is very helpful for SUSY scenarios with light, i.e., $\Order(100)\GeV$, SUSY particles.
In this section, focusing on this advantage, we will consider an application of the CP-safe gravity mediation model.

The anomalous magnetic moment of muons, or the muon $g-2$, has a $3\sigma$-level discrepancy between the experimental value measured in the Brookheaven E821 experiment~\cite{g-2_bnl2010} and theoretical predictions based on the Standard Model~\cite{g-2_davier2010,g-2_hagiwara2011}.
SUSY is capable to solve the anomaly with contributions of loop diagrams in which smuons, sneutrinos, neutralinos and charginos are involved~\cite{SUSY_gminus2}.
The SUSY contribution, which we call $\Delta a_\mu^{\rm SUSY}$, can be large enough to solve the discrepancy if the masses of bino and/or winos are $\lesssim1\TeV$ and those of smuons and/or muon sneutrinos are at the same order (see, e.g., Ref.~\cite{Chakraborti:2014gea} for recent study).
However, as we have discussed, this scenario generally confronts too large CP violation because of the small SUSY particle masses. This is the main reason why we apply the CP-safe gravity mediation model to explain the anomaly of the muon $g-2$.

\begin{figure}[t]
 \centering
 \includegraphics[width=.6\textwidth]{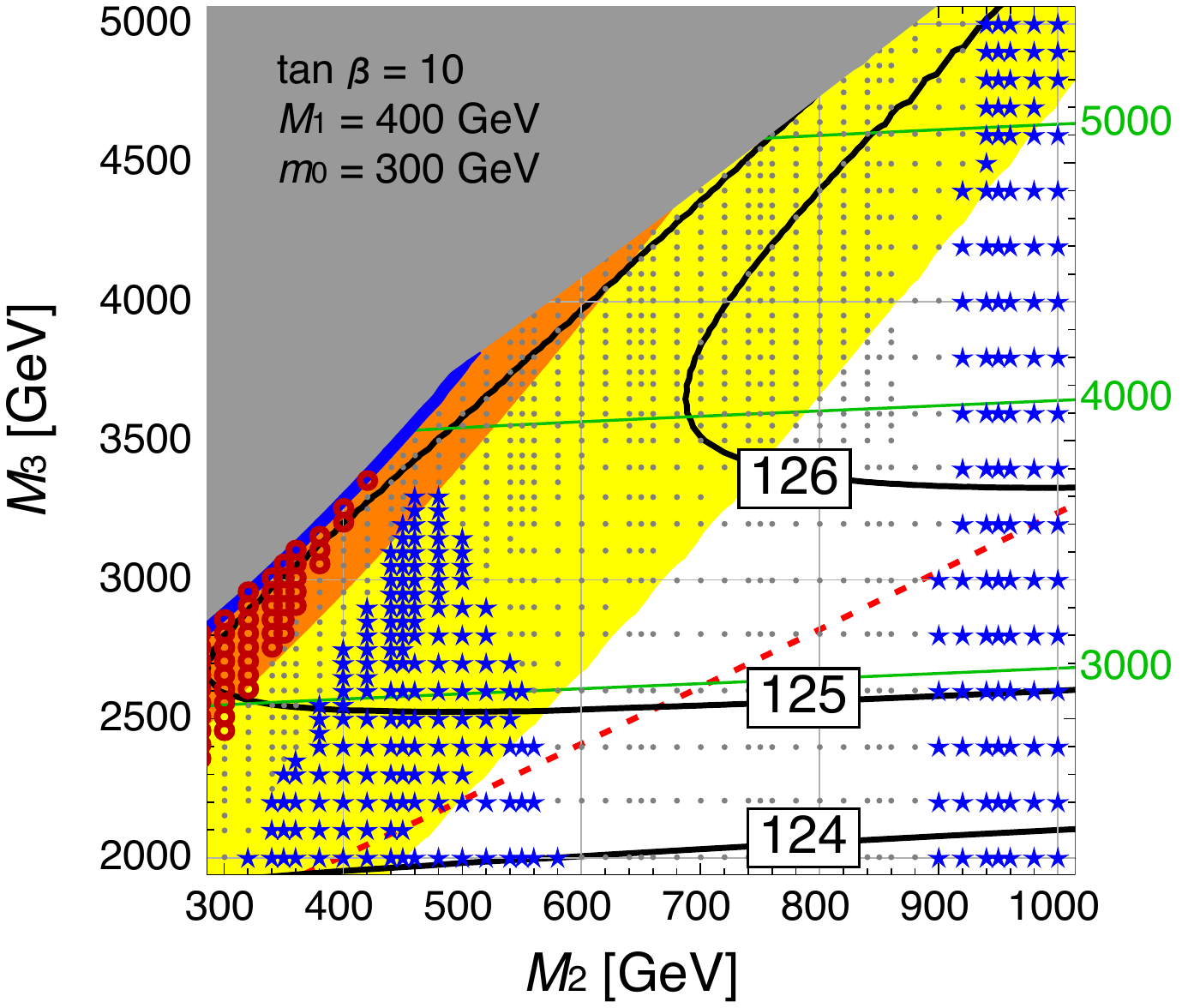}
 \caption{%
Higgs boson mass and the SUSY contribution to the muon $g-2$ in CP-safe gravity mediation.
The muon $g-2$ is compatible with the experimental value at $1\sigma$ level in the orange region and at $2\sigma$ level in the yellow.
The black contours show the Higgs boson mass, while the green contours describe the value of the $\mu$ parameter at the SUSY scale.
The gray-shaded region is invalid since the vacuum is unstable or the LSP is the lighter stau; the lighter stau is degenerated with the lightest neutralino on the blue line.
The plotted points correspond to the model point we analyzed. Red circles are the model points excluded by ATLAS slepton searches~\cite{Aad:2014vma}, while blue stars are those which we have checked is not excluded in any collider searches.
Collider status of the other points (gray dots) are not determined in this study.
The parameter space with $\alpha_1=\alpha_2=0$ is described with the red-dashed line as a reference.
}
\label{fig:m2_m3}
\end{figure}

In this section, we consider a model with slightly extended K\"ahler potential,
\begin{equation}
\begin{split}
   K &= s(x) + (Q_{\rm SM}^*)_i (Q_{\rm SM})_i \left[\frac{ 1 + \alpha_1 x + \alpha_2 x^2}{r}\right],
 \\
  W &= \mu H_u H_d + W_{\rm Yukawa}+ \mathcal{C},
\end{split}
\end{equation}
where $\alpha_1$ and $\alpha_2$ are real constants and $x=Z+Z^*$.
A normalization factor $r=1+\alpha_1\l<x\r> + \alpha_2\l<x\r>^2$ is introduced for canonical kinetic terms.
Note that $\alpha_1=\alpha_2=0$ corresponds to the model discussed in the previous section (Eq.~(\ref{eq:cpsafe_lag})), and that this extension introduces no additional CP phases.\footnote{The original model with $\alpha_1=\alpha_2=0$ has a strict constraint that the universal scalar mass parameter at the high-energy scale is related to the Higgs B-term as $(B_\mu/\mu)^2 = 4 m_0^2$ (cf.~Eqs.~(\ref{eq:b_and_a}) and (\ref{eq:scalar_soft_mass})).}
The Higgs B-term and A-terms are given by
\begin{equation}
\begin{split}
   B_{\mu} &= -\frac{2}{3}\l[ \l<\frac{\q s}{\q x}\r> - 3\alpha_1' M_P^2 \r] \frac{\l<F_Z\r>}{M_P^2} \mu ,
\\
 A_{ijk}&= - \l[\l<\frac{\q s}{\q x}\r> - 3\alpha_1' M_P^2 \r]  \frac{\l<F_Z\r>}{M_P^2},
\end{split}
\end{equation}
and the universal scalar mass is by
\begin{equation}
 m_0^2 =  \frac{|\l<F_Z\r>|^2}{M_P^4} \l[\frac{1}{9}\l<\frac{\q s}{\q x}\r>^2   + \l(\alpha_1'^2-\alpha_2'\r)M_P^4 \r],
\end{equation}
where $\alpha_1'\equiv (\alpha_1+2\alpha_2 \langle x\rangle)/r$ and $\alpha_2'\equiv 2\alpha_2/r$.

Now, the parameters of this model are summarized as $(m_0, \tan\beta, \sgn\mu, M_i)$;
$m_0$ and $M_i$ are the scalar and gaugino mass parameters at the high-energy scale, respectively, and $\tan\beta$ is the ratio of the VEVs of the Higgs fields: $\tan\beta=v\s u/v\s d$.
Note that trilinear couplings are universal and fixed as $A_{ijk}/(B_\mu/\mu) = 3/2$ at the high-energy scale even in this extended model.

A series of benchmark points are shown in Fig.~\ref{fig:m2_m3}.
The Higgs boson mass is plotted with black lines, while the muon $g-2$ discrepancy is relaxed to less than $1\sigma$ ($2\sigma$) level in the orange (yellow) region.
Green lines show the size of $\mu$ parameter at the SUSY scale.
The lightest neutralino is the lightest SUSY particle (LSP), and it is bino-like, in the displayed parameter space.
The gray-shaded region is invalid or excluded because of the stau LSP, existence of tachyonic SUSY particles, and/or EWSB vacuum instability~\cite{Hisano:2010re,Endo:2013lva}.
The lighter stau is degenerated with the lightest neutralino on the blue line in the figure, where the dark matter relic density can be explained by the coannihilation mechanism~\cite{stau_coann}.
The values of the fixed parameters are $(m_0,\tan\beta,\sgn\mu,M_1)=(300\GeV,10,+,400\GeV)$.
As a reference, the parameter space of the model discussed in Sec.~3, i.e., $\alpha_1=\alpha_2=0$, is described with the red-dashed line.
Mass spectra are calculated with {\tt SOFTSUSY\,3.4}~\cite{SOFTSUSY}. The mass of the Higgs boson and $\Delta a_\mu^{\rm SUSY}$ is obtained with {\tt FeynHiggs\,2.10}~\cite{FeynHiggs}, where the calculation of $\Delta a_\mu^{\rm SUSY}$ is restricted to the one-loop level.\footnote{See Ref.~\cite{SUSY_g-2_stockinger} for detailed discussion on two-loop level contributions in scenarios with hierarchical SUSY mass spectrum.}
{\tt SDECAY\,1.3}~\cite{Muhlleitner:2003vg} is also utilized to obtain decay rates and branching ratios of the SUSY particles.
For the analysis of the vacuum stability, we utilized the fitting function in Ref.~\cite{Endo:2013lva}.

When we fix $M_2$ and increase $M_3$, $\Delta a_\mu^{\rm SUSY}$ becomes larger and the discrepancy of the muon $g-2$ is relaxed, but with too large $M_3$ we face vacuum instability or the stau LSP.
This feature can be understood with the following discussion on the renormalization group evolution from a high-energy scale of $\sim10^{16}\GeV$ to the SUSY scale.
Firstly, the large $M_3$ increases squark masses during the evolution.
Then, the large scalar-top mass, which on the one hand raises the Higgs boson mass, affects the soft mass of the up-type Higgs.
For successful EWSB, the $\mu$ parameter is forced to be large, as shown in Fig.~\ref{fig:m2_m3}.
It results in a large mixing between $\tilde\mu\s L$ and $\tilde \mu\s R$, which enhances the bino--smuon contribution (a loop diagram of $\tilde B$--$\tilde \mu\s L$--$\tilde \mu\s R$).
Note that the other contributions, e.g., sneutrino--chargino contribution, are insignificant because Higgsinos are decoupled.
However, too large mixing is not allowed because it makes the lighter stau too light or induces vacuum instability~\cite{Hisano:2010re,Endo:2013lva}, and this is why $M_3$ is bounded from above.

Fig.~\ref{fig:m2_m3} also contains the information on the status of collider searches.
Because squarks and gluinos are much heavier than the current collider bounds~\cite{Aad:2014wea,CMSPASSUS13019}, LHC searches for electroweakino (sleptons\footnote{In this section we use the term ``lepton'' (``$l$'') for electrons and muons, not taus.}, staus, neutralinos, and charginos) gives the most severe constraint on our benchmark points.
However, in order to discuss the collider search status on all the model points, we have to perform dedicated Monte Carlo simulation, which is far beyond the scope of this paper.
We thus leave the status of collider searches unascertained for most of the points, and show the status for clearly excluded and clearly surviving points in Fig.~\ref{fig:m2_m3}.
The model points depicted with red circles are excluded by di-lepton plus large missing energy ($2l+\MET$) signature from $pp\to\tilde l\tilde l^*$~\cite{Aad:2014vma}, while those with blue stars are not excluded by any searches.
The other points, i.e., those with gray dots, are unascertained points.
Here, because there is no significant difference between results by CMS collaboration~\cite{Khachatryan:2014qwa,Chatrchyan:2014aea} and by ATLAS~\cite{Aad:2014nua,Aad:2014vma,Aad:2014iza}, the status is determined using the results from ATLAS collaboration.

\begin{figure}[t]
\centering
 \includegraphics[width=0.6\textwidth]{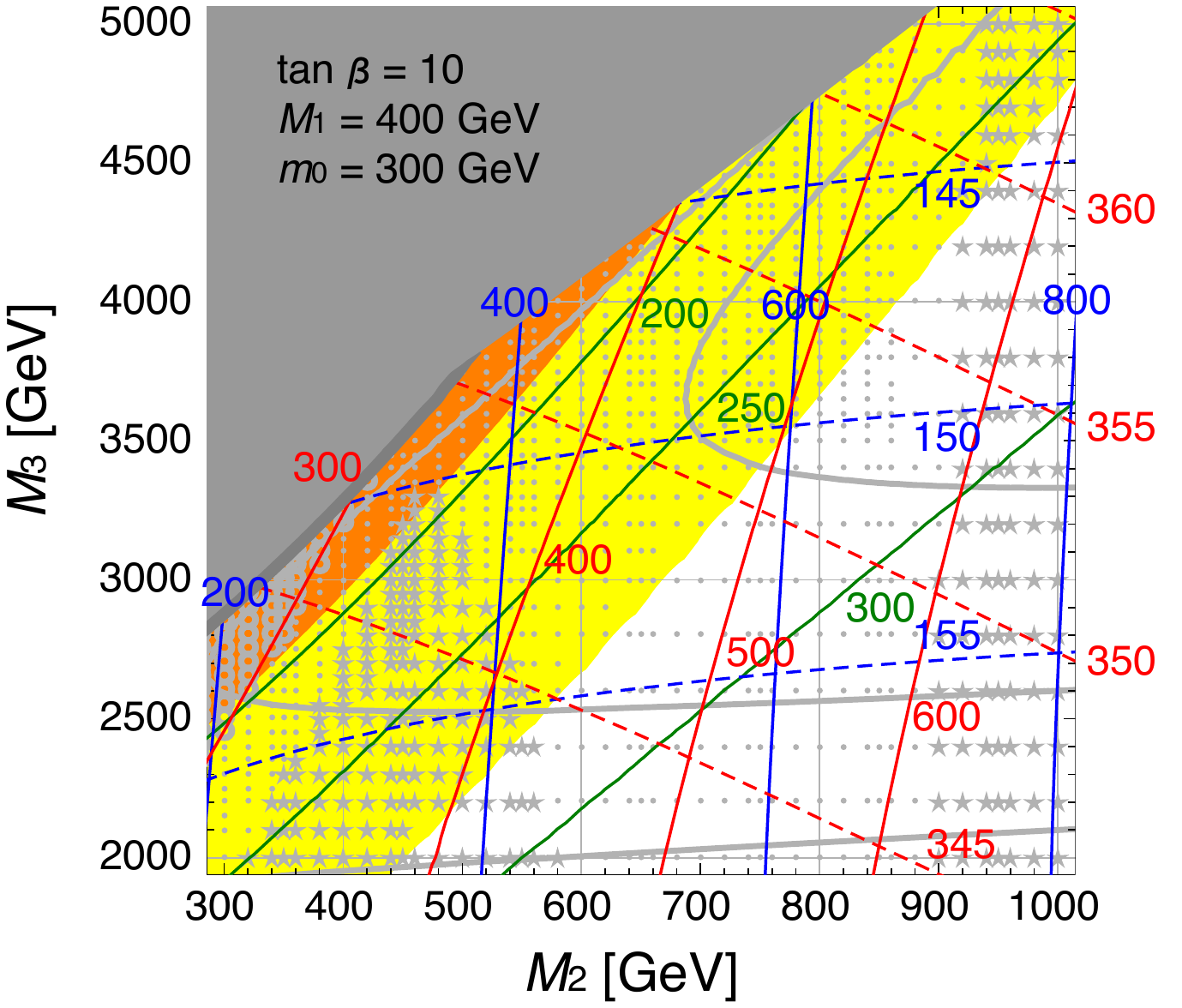}
 \caption{%
Masses of electroweakinos in the parameter space of Fig.~\ref{fig:m2_m3}.
Blue solid contours show the mass of winos (the lightest chargino $\tilde\chi^\pm_1$ and the second-lightest neutralino $\tilde\chi^0_2$), and blue-dashed contours are those of bino-like LSP $\tilde\chi^0_1$.
Red-solid (dashed) lines describe the masses of left- (right-) handed selectrons and smuons, and green solid lines do the masses of the lighter staus.}
 \label{fig:masses}
\end{figure}

To see details of the collider search status, we pick up two benchmark points among the surviving points and show the mass spectrum as well as the Higgs boson mass and $\Delta a_\mu^{\rm SUSY}$ in Table~\ref{tab:spectrum}.
The first point, $(M_2,M_3)=(460\GeV,3.3\TeV)$, explains the muon $g-2$ at $0.9\sigma$ level.
For this point, the sleptons are slightly heavier than the current bound~\cite{Aad:2014vma} ($m_{\tilde l}>324\GeV$ for $m_{\tilde\chi^0}=150\GeV$ if $m_{\tilde l_L}=m_{\tilde l_R}$).
Wino pair-production $pp\to\tilde\chi^0_2\tilde\chi^+_1$ is generally the most important for SUSY scenarios compatible with large $\Delta a_\mu^{\rm SUSY}$~\cite{Endo:2013bba} for its cross section of $\Order(10)\,{\rm fb}$. At this point, however, the winos mainly decay into staus and tau-sneutrinos to result in multi-tau signature,  which is less constrained than the signature with winos decaying into leptons ($e$ or $\mu$).
For the second point, $(M_2,M_3)=(1\TeV,5\TeV)$, the masses of electroweakinos are clearly above the collider bounds in Refs.~\cite{Aad:2014nua,Aad:2014vma,Aad:2014iza}.
Further information on the electroweakino masses can be found on Fig.~\ref{fig:masses}.

\begin{table*}[t]
 \centering
 \caption{Masses of SUSY particles, Higgs boson mass and SUSY contribution to the muon $g-2$ at two benchmark points. Masses are given in the unit of GeV. $M_2$ and $M_3$ are values at the high-energy scale in the unit of GeV. Values in the braces on $\Delta a_\mu^{\rm SUSY}$ are the deviation from the experimental results, where we use the value $(26.1\pm8.0)\times 10^{-10}$ as the discrepancy of the muon $g-2$~\cite{g-2_hagiwara2011}. Note that $\tilde\chi^0_2$ and $\tilde\chi^\pm_1$ are degenerated because they are mostly pure-wino.}
\label{tab:spectrum}
 \begin{tabular}[t]{ccccccccc}\toprule
 $(M_2,M_3)$        &
 $\tilde\chi^0_2,\tilde\chi^\pm_1$ &
 $\tilde l_L$       &
 $\tilde l_R$       &
 $\tilde\tau_1$     &
 $\tilde\tau_2$     &
 $\tilde\chi^0_1$   &
 $m_h$              &
 $\Delta a_\mu^{\rm SUSY}\times10^{10}$\\\midrule
 $(460,3300)$       & 333 & 329 & 348 & 182 & 400 & 150 & 125.3 & 18.8 ($0.9\sigma$) \\
 $(1000,5000)$      & 774 & 578 & 365 & 247 & 589 & 142 & 126.5 & 11.2 ($1.9\sigma$) \\\bottomrule
 \end{tabular}
\end{table*}

Let us discuss future prospects of collider searches for this scenario.
From the viewpoint of the future prospect, the CP-safe gravity mediation model as a solution to the muon $g-2$ anomaly has three important features.
The first one is the large $M_3$, which is indicated by the collider bound and the Higgs boson mass, and results in the large $\mu$ parameter and decoupled Higgsinos.
It diminishes the chargino contribution to the muon $g-2$.
Consequently, as the second point, both $\tilde\mu\s L$ and $\tilde\mu \s R$ are as light as $\Order(100)\GeV$ to yield sufficient contribution to the muon $g-2$.
Thirdly, the universal scalar soft masses \eqref{eq:scalar_soft_mass} ensure that the lighter stau is lighter than the smuons.

The most important signature is $2l+\MET$ from $pp\to\tilde l\tilde l^*$.
It is generally important for any SUSY scenarios explaining the muon $g-2$ anomaly but because of light $\tilde\mu\s L$ and $\tilde\mu\s R$ it is more effective for this scenario.
Because the cross section $\sigma(pp\to\tilde\mu\s L\tilde\mu^*\s L)$ at the $14\TeV$ LHC with $m_{\tilde \mu\s L}=400\GeV$ is approximately equal to that at $8\TeV$ with $m_{\tilde\mu\s L}=300\GeV$, whole of the $1\sigma$ region in Fig.~\ref{fig:m2_m3} will be examined at the $14\TeV$ LHC.

Another promising production channel is wino production $pp\to\tilde\chi^0_2\tilde\chi^+_1$, which has the largest production cross section.
Future prospects of this channel are widely discussed~\cite{ATLASPHYSPUB2013002,ATLASPHYSPUB2013011,CMS:2013xfa,CMSPASFTR13014} in $WZ+\MET$ signature, i.e., with assuming that produced winos exclusively decay as $\tilde\chi^0_2\tilde\chi^+_1\to (Z\tilde\chi^0_1)(W^+\tilde\chi^0_1)$.
Reportedly, winos with a mass of 700--800$\GeV$ are searched for at $14\TeV$ LHC with the integrated luminosity of $300\,{\rm fb}^{-1}$ under the assumption.\footnote{%
In Ref.~\cite{Baer:2012vr} future prospects of this channel is discussed in $WH+\MET$ signature.}
However, the branching ratio is negligible in our benchmark points; the winos mostly decays via sleptons, sneutrinos or staus, and produces multi-$(e,\mu,\tau)$ plus $\MET$ signature.
As capability of searches for these signature seriously dependent on mass gaps among SUSY particles as well as resultant lepton species~\cite{Endo:2013bba}, we will here just comment that $\text{multi-}l+\MET$ signature generally provides tighter constraint than $WZ+\MET$ while $\text{multi-}\tau+\MET$ gives looser bound~\cite{Chatrchyan:2014aea,Aad:2014nua}.

For experiments at linear colliders, e.g., ILC or CLIC, the stau is a particularly interesting target for its small mass and large mixing.
After staus are discovered, stau mixing angle as well as stau mass should be measured, because it will be a test of this scenario.
Furthermore, if all the sleptons are within the reach of linear colliders, we can estimate the size of $\Delta a_\mu^{\rm SUSY}$ through measurements of the stau mixing angle, slepton production cross section and masses~\cite{Endo:2013xka}.


%
%
%
%
\section{Discussion and conclusions}

We have discussed the CP-safe gravity mediation models, which are free from the SUSY CP problem. 
The SUSY CP problem is serious obstacle for the low-energy SUSY.
The CP-safe gravity mediation models assume that SUSY is broken in the K\"ahler potential with a non-vanishing constant term in the superpotential. Together with a shift symmetry of the SUSY breaking field, the Higgs B-term, trilinear couplings as well as gaugino masses have a common phase determined by the constant term in the superpotential, and hence, their phases are all aligned. 

Note that the CP-safe gravity mediation is also consistent with the sequestered form of the K\"ahler potential. (A concrete model is shown in Appendix D.) In this case, the Higgs B-term, A-terms and scalar masses vanish at the tree level; then,  the gaugino mediation with a large gravitino mass is possible~\cite{focuspoint_gauginomed},\footnote{
For a large gravitino mass of $\Order(100)$ TeV, the coupling constants $k_i (i=1-3)$ in Eq.~(\ref{eq:z_gauge}) may be taken small as $k_i = \Order(0.01)$ to produce $\Order(1)$ TeV gaugino masses. Such small $k_i$ seem more likely if the terms in Eq.~(\ref{eq:z_gauge}) are generated from gauge anomalies of the shift symmetry.
}
and the gravitino problem may be relaxed. It is very important that the stable EWSB vacuum is easily realized, since the Higgs B-term ($B_\mu/\mu$) becomes order of gaugino masses rather than $\Order(m_{3/2})$.
This is a clear advantage over the framework with $Z$-dependent superpotential, where the Higgs B-term of $\Order(m_{3/2})$ arises (see discussion in Appendix D).


In the CP-safe gravity mediation, the Lagrangian has the shift symmetry of the SUSY breaking field $Z$: $Z \to Z + i \mathcal{R}$ ($\mathcal{R}$ is a real constant). Therefore, the imaginary part of $Z$ is massless at the perturbative level. This imaginary part gets a mass via the QCD non-perturbative effect, since it couples to the gluon field strength superfield  as shown in Eq.~(\ref{eq:z_gauge});\footnote{
One might introduce a shift symmetry breaking term, e.g., $K \ni \epsilon_2 |Z|^2$ ($K \ni \epsilon_4 |Z|^4$).
With this shift breaking term, the axionic part of $Z$ can get a larger mass of $\sim {\epsilon_2} m_{3/2}$ ($\sim \sqrt{|\epsilon_4|} m_{3/2}$) without inducing a CP violating phase. However, such explicit breaking terms must be extremely small, otherwise the Peccei--Quinn mechanism does not work for solving the strong CP problem in QCD.
}
the imaginary part of $Z$ behaves as a QCD axion~\cite{pq_mechanism} with the decay constant,
\begin{equation}
 f_a = \frac{M_P}{k_3 16 \sqrt{2} \pi^2} \l(\frac{\q^2 s}{\q x^2}\r)^{1/2} \sim 10^{16} {\rm GeV}/k_3.
\end{equation}
It is very intriguing that the strong CP problem and the SUSY CP problem are solved simultaneously in the present framework. If the inflation scale is high as in the chaotic inflation, we have an axion iso-curvature problem. It may be solved in more complicated frameworks~\cite{Evans:2014hda}, which is, however, beyond the scope of the present paper.

\section*{Acknowledgment}
We thank K. Harigaya for helpful discussion.
This work was supported by JSPS KAKENHI Grant No.~26287039 and No.~26104009 (T.T.Y), and also by World Premier International Research Center Initiative (WPI Initiative), MEXT, Japan. The works of S.I. and N.Y. are supported in part by JSPS Research Fellowships for Young Scientists.

\appendix

\section{General superpotential}

The scalar potential is given by
\begin{equation}
 V = F_i^* K^{ij} F_j -3e^K|W|^2,  
\end{equation}
where 
\begin{align}
  K^{ij} &= \frac{\q^2 K}{\q \phi_i^* \q \phi_j},
&
 F_i^* &= -e^{K/2} (K^{-1})_{il} (\frac{\q W}{\q \phi_l} + \frac{\q K}{\q \phi_l} W).
\end{align}
We take the unit of $M_P=1$.
The gravitino mass is $m_{3/2}^2 = (F_i^* K^{ij} F_j)/3 \simeq e^{K} |W|^2$.
Assuming the minimal K\"ahler potential, $K=Z^* Z + (Q_{\rm SM})_i^{*} (Q_{\rm SM})_i$, 
the scalar potential is written as
\begin{equation}
  e^{-|z|^2/2}V_{\rm vis} 
\ni  \frac{\q W_{\rm vis}}{\q \phi_{\rm SM}^i} \phi_{\rm SM}^i (\left<W_{\rm hid}\right>^* +\mathcal{C}^*)
  + \l[\frac{\q W_{\rm hid}^*}{\q Z^*} z^* + (|z|^2 -3) (\left<W_{\rm hid}\right>^* + \mathcal{C}^*)\r] W_{\rm vis}
  + \text{h.c.}
\end{equation}
for the superpotential $W=W\s{vis}+W\s{hid}+\mathcal C$, where $z=\left<Z\right>$ and $W_{\rm vis}= \mu H_u H_d + W_{\rm Yukawa}$. The Higgs B-term and the A parameter are given by
\begin{equation}
 \begin{split}
  B_\mu/\mu &= \Biggl[\left(z+\frac{1}{W^*_{\rm hid}+\mathcal{C}^*}\frac{\q W_{\rm hid}^*}{\q Z^*}\right)^{-1} - z^* \Biggr] \l<F_Z\r>,
\\
A_{ijk} &= -z^* \l<F_Z\r>,
 \end{split}
\end{equation}
where 
$$
\l<F_Z\r> = - e^{|z|^2/2} \l[\frac{\q W_{\rm hid}^*}{\q Z^*} + \frac{\q K}{\q Z^*} (W_{\rm hid}^*+\mathcal{C}^*)\r].
$$
The gaugino mass arises from the coupling between $Z$ and the gauge kinetic function. 
\begin{equation}
 \int d^2 \theta \left(\frac{1}{4g_i^2} + k_i Z\right) W^{i \alpha} W_{\alpha}^i  + \text{h.c.}
\end{equation}
The resultant gaugino masses are $M_i = 2 k_i g_i^2 F_Z$.

%
%
%
%
%
%
%
%
%
%
%
%
\section{The Gravitational SUSY breaking}
Here, we consider a model where SUSY is broken with a constant superpotential.
The SUGRA Lagrangian is written as
\begin{equation}
 \mathcal{L} = \int d^4 \theta \varphi \varphi^* f + \int d^2 \theta \varphi^3 (\mathcal{C} + W_{\rm vis}) + \text{h.c.},
\end{equation}
where $\varphi$ is the chiral compensator and $\varphi=1+F\theta^2$. Here, $W_{\rm vis}$ is a function of the SSM fields.
Then, the scalar potential is given by
\begin{equation}
 \begin{split}
  -V &= |F|^2 f + F^* F_i \frac{\q f}{\q Q_i} + F_i^* F \frac{\q f}{\q Q_i^*} + \frac{\q^2 f}{\q Q_i^* \q Q_j} F_i^* F_j
\\&
+ 3(\mathcal{C} + W_{\rm vis})F + 3(\mathcal{C}^* + W_{\rm vis}^*)F^*
 + \frac{\q W_{\rm vis}}{\q Q_i} F_i + \frac{\q W_{\rm vis}^*}{\q Q_i^*} F^*_i,
 \end{split}
\end{equation}
where $Q_i$ contains both the SUSY breaking field $Z$ and the SSM fields.
The equations of motions are 
\begin{align}
  &F f + F_i \frac{\q f}{\q Q_i} + 3(\mathcal{C}^* + W_{\rm vis}^*)=0,
 &
 &F \frac{\q f}{\q Q_k^*} + \frac{\q^2 f}{\q Q_k^* \q Q_j} F_j + \frac{\q W_{\rm vis}^*}{\q Q_k^*} = 0. \label{eq:eom_grav}
\end{align}
Solving equation of motions, $F_i$ is written as
\begin{equation}
 F_i = -(\tilde{f}^{-1})_{ik} \l[ 
\frac{\q W_{\rm vis}^*}{\q Q_k^*} - \frac{1}{f} \frac{\q f}{\q Q_k^*} 3 (\mathcal{C}^*+W_{\rm vis}^*)
\r],
\end{equation}
where $\tilde{f}^{-1}$ is the inverse of the matrix
\begin{equation}
 \tilde{f}^{ij} = \frac{\q^2 f}{\q Q_i^* Q_j} -\frac{1}{f} \frac{\q f}{\q Q_i^*} \frac{\q f}{\q Q_j}.
\end{equation}
The scalar potential is 
\begin{align}
 V&=-3(\mathcal{C} + W_{\rm vis})F - \frac{\q W_{\rm vis}}{\q Q_i} F_i
\nn
\begin{split}
 &=
 \frac{9|\mathcal{C}+W_{\rm vis}|^2}{f}
 + \left[ 3(\mathcal{C} + W_{\rm vis}) \frac{1}{f} \frac{\q f}{\q Q_i} - \frac{\q W_{\rm vis}}{\q Q_i}\right] (\tilde{f}^{-1})_{ij}
 \left[ 3(\mathcal{C}^* + W_{\rm vis}^*) \frac{1}{f} \frac{\q f}{\q Q_j^*} - \frac{\q W_{\rm vis}^*}{\q Q_j^*}\right].
\end{split}
\end{align}
The scalar potential in the Einstein frame is obtained by the rescaling, $V \to (-3/f)^2 V$.
The vanishing cosmological constant is obtained by requiring $V=0$, resulting in
\begin{equation}
 \begin{split}
   -\l< F \r> 
&=
 \frac{1}{f} \left(F_Z \frac{\q f}{\q Z} + 3 \mathcal{C}^*\right)
\\& =
\frac{3\mathcal{C}^*}{f} \left[ \frac{1}{f} \l|\frac{\q f}{\q Z}\r|^2 \left(\frac{\q^2 f}{\q Z \q Z^*} - \frac{1}{f} \l|\frac{\q f}{\q Z}\r|^2 \right)^{-1}+1 \right]
\\& = 0.
 \end{split}
\end{equation}

Then, the K\"ahler potential $K=-3 \log (-f/3)$ must satisfy the condition:
\begin{equation}
 \l| \frac{\q K}{\q Z} \r|^2 \l(\frac{\q^2 K}{\q Z \q Z^*} \r)^{-1} \Bigg|_{Z=\l<Z\r>}= 3.
\end{equation}
To satisfy the condition for the vanishing cosmological constant, SUSY must be broken when $\mathcal{C} \neq 0$~\cite{Izawa:2010ym}. This is because $\l< F_i (\q f)/(\q Q_i) \r>$ can not be zero for $\l< F \r>=0$  (see Eq.~(\ref{eq:eom_grav})). 

Since $\l<F\r>=0$, the Higgs B-term and the A-terms arise from $V \ni - (\q W_{\rm vis}) / (\q Q_i) F_i$. The scalar masses arise from $V \ni - 3 \mathcal{C} F \ni 3 \mathcal{C} (1/f)(\q f/\q Q_i) F_i$.

\section{A model of CP-safe gravity mediation}

We consider the following K\"ahler potential of the hidden sector;
\begin{equation}
 K = s(x) = -3 \log (-f(x)/3),  
\end{equation}
where, $f$ is a function of $x= Z+Z^\dag $ and is invariant under the shift $Z \to Z + i \mathcal{R}$ with a real constant $\mathcal R$.
The superpotential is taken as $W=\mathcal{C}$; SUSY is broken with a constant superpotential.

By using the equations of motions, the scalar potential is written as $V = -3 \mathcal{C}F$, where
\begin{equation}
 -\l< F \r> = 
\frac{3c^*}{f} \left[ \frac{1}{f} \l|\frac{\q f}{\q x}\r|^2 \left(\frac{\q^2 f}{\q x^2} - \frac{1}{f} \l|\frac{\q f}{\q x}\r|^2 \right)^{-1}+1 \right] =0.
\end{equation}

The vanishing cosmological constant is given by $\left<F\right>=0$, which leads to
\begin{equation}
 \frac{\partial^2 f}{\partial x^2} \Bigg|_{x = \left< x \right>}=0.
\end{equation}
The minimum of $x$ is determined by $(\q V)/(\q x)=0$;
\begin{equation} 
\frac{\partial^3 f}{\partial x^3} \Bigg|_{x=\l<x\r>}= 0.
\end{equation}
The F-term of $Z$ is 
\begin{equation}
 \left<F_Z\right> =  (\tilde{f}^{-1})_{ZZ} \left( \frac{1}{f} \frac{\q f}{\q x} 3 \mathcal{C}^*\right) = -3 \l(\frac{\q f}{\q x}\r)^{-1} \mathcal{C}^*.
\end{equation}

Now, let us consider the following $f$; $f=-3 + c_1 x + c_2 x^2 + c_3 x^3 + c_4 x^4$. 
The conditions $\frac{\q f}{\q x} \neq 0$, $\frac{\q f^2}{\q x^2} =0$, and $\frac{\q^3 f}{\q x^3} =0 $ give
\begin{align}
 & \left<x\right>= -c_3/(4 c_4),&
 & c_2 + 3\left<x\right> (c_3 + 2 c_4 \left<x \right>)=0 \to c_2 = (3/8) c_3^2/c_4 .
\end{align}
The SUSY breaking vacuum is stable for $768c_4^3+64c_1 c_3 c_4^2 -3c_3^4 > 0$.

\section{A model with a sequestered K\"ahler potential}
Here, we consider the sequestered form of the K\"ahler potential and superpotential;
\begin{equation}
 f = -3  + f_{\rm vis} + f_{\rm hid}, \ \ K = -3 \log (-f/3),  \ \ W=\mathcal{C}+W_{\rm vis},
\end{equation}
where, $f_{\rm hid}$ is a function of $x= Z+Z^\dag $; $f$  is invariant under the shift, $Z \to Z + i \mathcal{R}$. 
The superpotential in the visible sector is $W_{\rm vis}= (\mu_{ij}/2) Q_i Q_j + (y_{ijk}/6) Q_i Q_j Q_k$.

If one choose $f_{\rm vis}=(Q_{\rm SM})_i^* (Q_{\rm SM})_i$, the Higgs B-term, A-terms, and scalar masses vanish. This can be seen clearly in the base that the SM fields are rescaled as $(Q_{\rm SM})_i \to (Q_{\rm SM})_i \varphi$. Then, the scalar potential is written as
\begin{equation}
\begin{split}
 -V &= f_{\rm hid}' |F|^2 +  (\frac{\q f_{\rm hid}'}{\q x} F_Z F^* + \text{h.c.})
 \\ &
 + \frac{\q^2 f_{\rm hid}'}{\q x^2} |F_Z|^2 + F_i^* F_i
 + (3\mathcal{C} F+ \frac{\mu_{ij}}{2} Q_i Q_j F + \text{h.c.}) + (\frac{\q W_{\rm vis}}{\q Q_i} F_i + \text{h.c.}) ,
\end{split}\end{equation}
where $f_{\rm hid}' = -3 + f_{\rm hid}$. The equations of motions are given by
\begin{equation}
\begin{split}
  f'_{\rm hid} F + \frac{\q f_{\rm hid}'}{\q x} F_Z + (3 \mathcal{C}^* + \frac{\mu_{ij}^*}{2} Q_i^* Q_j^*) &= 0,
\\
 \frac{\q f'_{\rm hid}}{\q x} F + \frac{\q^2 f_{\rm hid}'}{\q x^2} F_Z &= 0,\\
 F_i + \frac{\q W_{\rm vis}^*}{\q Q_i^*} &= 0.
\end{split}\end{equation}
Using equations of motions, we have
\begin{equation}
 V = -(3 \mathcal{C} +  \frac{\mu_{ij}}{2}Q_i Q_j)F - \frac{\q W_{\rm vis}}{\q Q_i} F_i,
\end{equation}
where $F_i=-({\q W_{\rm vis}^*})/({\q Q_i^*})$. Since the vanishing cosmological constant is satisfied with $\l<F\r>=0$~\cite{Izawa:2010ym},  B-terms and A-terms as well as scalar masses vanish at the tree level.

This vanishing B-term is an important advantage of our setup, because it allows us to easily realize the EWSB even if the gravitino is as heavy as $\mathcal{O}(10)\TeV$.
This may be very useful for gaugino mediation models~\cite{focuspoint_gauginomed}.
We can see this advantage by observing the EWSB conditions
\begin{align}
 &\frac{m_Z^2}{2} = -|\mu|^2 -\frac{m_{H_u}^2 \tan^2\beta -  m_{H_d}^2 }{\tan^2\beta-1},
\label{eq:ewsb1}
 \\
  &B_\mu (\tan\beta + \cot\beta)= 2|\mu|^2 + m_{H_u}^2 +  m_{H_d}^2.
\label{eq:ewsb2}
\end{align}
If typical mass scale of SUSY particles is $m\s{soft}=\Order(1)\TeV$, $B_\mu/\mu$ must be as small as $\lesssim\Order(1)\TeV$.
In our setup, this condition is fulfilled easily.

By contrast, if the superpotential depends on $Z$, i.e., $({\partial W})/({\partial Z}) \neq 0$, the Higgs B-term $B_\mu/\mu=\mathcal{O}(m_{3/2})$ arises from the VEV of the compensator field, and it is difficult to satisfy the EWSB conditions with $m_{3/2}=\Order(10)\TeV$ and $m_{\rm soft}=\Order(1)\TeV$.
To see this, we consider $Z$-dependent superpotential
\begin{equation}
 W = {\mathcal C}+W\s{hid}(Z)+W\s{vis}.
\end{equation}
The Lagrangian is, with rescaling of $Q_i \varphi \to Q_i$,
\begin{equation}
  \mathcal{L} = \int d^4 \theta (f_{\rm hid}' |\varphi|^2 + Q_i^\dag Q_i) 
+ \Bigl\{
 \int d^2 \theta\Bigl[ (\mathcal C+W_{\rm hid})\varphi^3 
 \quad +\frac{\mu_{ij}}{2} \varphi Q_i Q_j + \frac{\lambda_{ijk}}{6} Q_i Q_j Q_k \Bigr]+\text{h.c.}\Bigr\},
\end{equation}
The equations of motions are given by
\begin{equation}
\begin{split}
  f'_{\rm hid} F + \frac{\q f_{\rm hid}'}{\q x} F_Z + (3 \mathcal{C}^* + 3W_{\rm hid}^* + \frac{\mu_{ij}^*}{2} Q_i^* Q_j^*) &= 0,
\\
 \frac{\q f'_{\rm hid}}{\q x} F + \frac{\q^2 f_{\rm hid}'}{\q x^2} F_Z + \frac{\q W_{\rm hid}^*}{\q Z^*}&= 0,\\
 F_i + \frac{\q W_{\rm vis}^*}{\q Q_i^*} &= 0.
  \end{split}\label{eq:ZdependentEOM}
\end{equation}
Using the above equations, the scalar potential is now written in a simple form:
\begin{equation*}
 -V=  (3\mathcal C+3W_{\rm hid}+\frac{\mu_{ij}}{2}Q_iQ_j)F + \frac{\partial W_{\rm hid} }{\partial Z} F_Z+ \frac{\partial W_{\rm vis}}{\partial Q_i} F_{i}.
\end{equation*}
The condition for vanishing cosmological constant, $V=0$, is read as
\begin{equation*}
 3(\mathcal C+W_{\rm hid})F + \frac{\partial W_{\rm hid} }{\partial Z} F_Z =0
\end{equation*}
at the vacuum. Because of non-zero $W\s{hid}$, $\left<F\right>$ is not zero but $\Order(m_{3/2})$, and it induces the Higgs B-term as $(B_\mu/\mu)\sim m_{3/2}$.%
\footnote{
%
Even with $Z$-dependent superpotential, it is possible to suppress the CP violating phase.
An example is the case that the arguments of $\mathcal{C}$, $\left<W\s{hid}\right>$ and $\left<\partial W\s{hid}/\partial Z\right>$ can be taken as the same by shift of $Z$, i.e., $Z \to Z + i \theta$. (For instance, $W\s{hid} = A e^{-bZ}$ with $b\in\mathbb R$.)
Then we observe $\arg(\left<F\right>) = \arg(\left<F_Z\right>)$ from the equations of motions \eqref{eq:ZdependentEOM}, and the B-term is free from CP violating phases.
}
This B-term is problematic for the condition \eqref{eq:ewsb2} when we consider scenarios with $m_{3/2}=\Order(10)\TeV$ and $(\mu, m_{H_u}, m_{H_d}) = O(0.1\text{--}1)\TeV$.%

{\small
\bibliography{CPsafe}
}
\end{document}